\documentclass[a4paper]{jpconf}
\usepackage{graphicx}
\usepackage{wrapfloat}
\usepackage{cite}

\begin{document}

\def\lsim{\,$\raise0.3ex\hbox{$<$}\llap{\lower0.8ex\hbox{$\sim$}}$\,}

\title{Frustrated \mbox{\boldmath{$S\!=\!1/2$}} Two-Leg Ladder with
Different Leg Interactions}

\author{
Takashi~Tonegawa$^{1,2}$, Kiyomi~Okamoto$^{3}$, Toshiya~Hikihara$^{4}$ and
T{\^o}ru~Sakai$^{5,6}$
}

\address{
$^{1}$Professor Emeritus, Kobe University, Kobe 657-8501, Japan\\
$^{2}$Department of Physical Science, Osaka Prefecture University,
Sakai, 599-8531, Japan\\
$^{3}$College of Engineering, Shibaura Institute of Technology,
Saitama, 337-8570, Japan\\
$^{4}$Faculty of Science and Technology, Gunma University, Kiryu,
376-8515, Japan\\
$^{5}$Graduate School of Material Science, University of Hyogo, Hyogo
678-1297, Japan\\
$^{6}$National Institutes for Quantum and Radiological Science and Technology
(QST), SPring-8, Hyogo 679-5148, Japan
}

\ead{tone0115@vivid.ocn.ne.jp}

\begin{abstract}
We explore the ground-state phase diagram of the \hbox{$S\!=\!1/2$} two-leg
ladder.  The isotropic leg interactions $J_{{\rm l},a}$ and $J_{{\rm l},b}$
between nearest neighbor spins in the legs $a$ and $b$, respectively, are
different from each other.  The $xy$ and $z$ components of the uniform rung
interactions are denoted by $J_{{\rm r}}$ and $\Delta J_{{\rm r}}$,
respectively, where $\Delta$ is the $XXZ$ anisotropy parameter.  This system
has a frustration when \hbox{$J_{{\rm l},a} J_{{\rm l},b}\!<\!0$} irrespective
of the sign of $J_{\rm r}$.  The phase diagrams on the
$\Delta$ (\hbox{$0\!\leq\!\Delta\!<\!1$}) versus $J_{{\rm l},b}$ plane
in the cases of \hbox{$J_{{\rm l},a}\!=\!-0.2$} and
\hbox{$J_{{\rm l},a}\!=\!0.2$}
with \hbox{$J_{{\rm r}}\!=\!-1$} are determined numerically.  We employ
the physical consideration, the level spectroscopy analysis of the results
obtained by the exact diagonalization method and also the density-matrix
renormalization-group method.  It is found that the non-collinear
ferrimagnetic (NCFR) state appears as
the ground state in the frustrated region of the parameters.  Furthermore, the
direct-product triplet-dimer (TD) state in which all rungs form the TD pair is
the {\it exact ground state}, when
\hbox{$J_{{\rm l},a}\!+\!J_{{\rm l},b}\!=\!0$} and
\hbox{$0\!\leq\!\Delta\lsim 0.83$}.  The obtained phase diagrams consist of
the TD, $XY$ and Haldane phases as well as the NCFR phase.
\end{abstract}

\section{Introduction}

In the past years a great deal of work has been devoted to the study which aims
at clarifying the role of the frustration in low-dimensional quantum spin
systems with competing interactions.  As regards the \hbox{$S\!=\!1/2$} two-leg
ladder systems, the general cases where additional leg next-nearest-neighbor
and/or diagonal interactions are competing with the leg nearest-neighbor and
rung interactions have been extensively
investigated~\cite{frustrated-leg-ladder1,frustrated-leg-ladder2,
frustrated-leg-ladder3}.  Very recently, we~\cite{ladder-ral-tone} have
discussed the ground-state phase diagram of the frustrated \hbox{$S\!=\!1/2$}
two-leg ladder, in which rung interactions are
ferromagnetically-antiferromagnetically alternating and have a common
Ising-type anisotropy, while leg interactions are antiferromagnetically uniform
and isotropic.  The phase diagram which we have numerically determined in the
case where the leg interactions are relatively weak compared with the rung
interactions shows that the incommensurate Haldane state as well as the
commensurate one appears as the ground state in the whole range of the
Ising-type anisotropy parameter.  This appearance of the Haldane state in the
case where the Ising character of rung interactions is strong is contrary to
the ordinary situation, and is called the inversion phenomenon concerning the
interaction anisotropy~\cite{inv-1,inv-2,inv-3,inv-4}.  The ground-state phase
diagram of the frustrated rung-alternating \hbox{$S\!=\!1/2$} two-leg ladder
in which all interactions are isotropic has also been studied by combining
analytical approaches with numerical simulations~\cite{ladder-ral-amiri}.
Furthermore, it has been shown that the introduction of the rung alternation
gives rise to the half-magnetization plateau in the ground-sate magnetization
curve~\cite{ladder-ral-japa}.  This result is consistent with the necessary
condition for the appearance of the magnetization plateau by Oshikawa, Yamanaka
and Affleck~\cite{OYA}.

\begin{figure}[htp]
  \begin{center}
    \scalebox{0.6}{\includegraphics{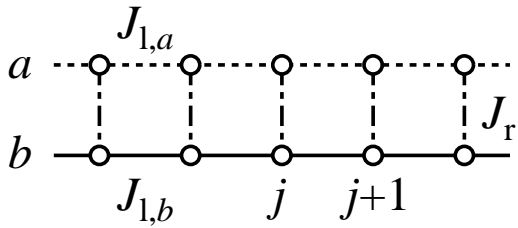}}~~~~~
    \begin{minipage}[b]{20pc}
       \caption{Sketch of the present model.
                Open circles denote \hbox{$S\!=\!1/2$} spins, and lines
                three kinds of interactions between spins.
                \bigskip\bigskip}
       \label{fig:model}         
    \end{minipage}
  \end{center}
\end{figure}

In the present paper, we explore the ground-state phase diagram of another
frustrated \hbox{$S\!=\!1/2$} two-leg ladder with different leg
interactions.  We express the Hamiltonian which describes this system as
\begin{equation}
{\cal H}
   = J_{{\rm l},a} \sum_{j=1}^{L}
                 {\vec S}_{j,a}\cdot{\vec S}_{j+1,a}
    + J_{{\rm l},b} \sum_{j=1}^{L}
                 {\vec S}_{j,b}\cdot{\vec S}_{j+1,b}
    + J_{\rm r} \sum_{j=1}^{L}\bigl\{
         S_{j,a}^x S_{j,b}^x\!+\!S_{j,a}^y S_{j,b}^y +
                                     \Delta S_{j,a}^z S_{j,b}^z\bigr\} \,.
\label{eq:hamiltonian}
\end{equation}
Here, ${\vec S}_{j,\ell}\!=\!\bigl(S_{j,\ell}^x,\,S_{j,\ell}^y,\,
S_{j,\ell}^z\bigr)$ is the \hbox{$S\!=\!1/2$} operator acting at the
($j$,$\,\ell$) site assigned by rung $j$ and leg $\ell(=\!a~{\rm or}~b)$;
$J_{{\rm l},a}$ and $J_{{\rm l},b}$ denote, respectively, the magnitudes of
the isotropic leg $a$ and leg $b$ interactions; $J_{\rm r}$ denotes that of
the anisotropic rung interaction, the $XXZ$-type anisotropy being
controlled by the parameter $\Delta$; $L$ is the total number of rungs, which
is assumed to be even.  The sketch of the present model is given in
Fig.~\ref{fig:model}.  It should be noted that this system has a frustration
when \hbox{$J_{{\rm l},a} J_{{\rm l},b}\!<\!0$} irrespective of the sign of
$J_{\rm r}$.

The most characteristic feature of the present system is the fact that, when
the condition \hbox{$J_{{\rm l},a}\!+\!J_{{\rm l},b}\!=\!0$}, which belongs to
the frustration region, is satisfied, the following three states are the
{\it exact eigenstates} of the Hamiltonian~(\ref{eq:hamiltonian}).

\vspace{-0.28cm}

\begin{itemize}
\parsep=0pt
\itemsep=0pt
\parskip=0pt
\item[1)] The direct-product singlet-dimer (SD) state in which all rungs form
the SD $\bigl((\alpha_{j,a}\beta_{j,b}-\beta_{j,a}\alpha_{j,b})/{\sqrt{2}}
\bigr)$ pair.

\item[2)] The direct-product triplet-dimer (TD) state in which all rungs form
the TD $\bigl((\alpha_{j,a}\beta_{j,b}+\beta_{j,a}\alpha_{j,b})/{\sqrt{2}}
\bigr)$ pair.

\item[3)] The nematic state with an arbitrary phase $\phi$ in which all rungs
are in the state given by a linear combination of two ferromagnetic states,
\hbox{$\cos\phi\,\alpha_{j,a}\alpha_{j,b}\!+\!\sin\phi\,\beta_{j,a}
\beta_{j,b}$}.

\end{itemize}

\vspace{-0.3cm}

\noindent
Here, $\alpha_{j,l}$ denotes the \hbox{$S_{j,l}^z\!=\!+1/2$} state and
$\beta_{j,l}$ the \hbox{$S_{j,l}^z\!=\!-1/2$} state.  These facts can be proven
by operating the Hamiltonian~(1) directly to the above three states.
Furthermore, it can be analytically shown that, when
\hbox{$J_{{\rm l},a}\!+\!J_{{\rm l},b}\!=\!0$},
\hbox{$J_{\rm r}\!<\!0$}, and the $XY$-type anisotropy of rung interactions
is sufficiently strong
$\Bigl($\hbox{$1\!-\!\frac{2|J_{{\rm l},a}|}{|J_{\rm r}|}\!\gg\!\Delta\!
\geq\!0$}$\Bigr)$, the direct-product TD state is the {\it exact ground state}
of the system, and that, when \hbox{$J_{{\rm l},a}\!+\!J_{{\rm l},b}\!=\!0$}
and \hbox{$J_{\rm r}(>\!0)$} is sufficiently large, the direct-product SD
state is the {\it exact ground state} of the system.  It is
noted that the above results concerning with the direct-product SD state has
already been shown by Tsukano and Takahshi~\cite{tsukano-takahashi}.  We also
note that all of the above results including the nematic state with $\phi$ as
well as the direct-product TD and SD states are applicable to systems in higher
dimensions, in which units of two \hbox{$S\!=\!1/2$} spins form
lattices; the details will be discussed in our forthcoming paper~\cite{HTOS}.

Unfortunately, materials corresponding to the present model have been neither
yet found nor synthesized so far.  We believe, however, that it is a physically
realistic model.  In fact, for example, Yamaguchi {\it et}
{\it al.}~\cite{yamaguchi-etal-1, yamaguchi-etal-2} have recently demonstrated
the modulation of magnetic interactions in spin ladder systems by using
verdazyl-radical crystals.  It is highly expected that the flexibility of 
molecular arrangements in such organic-radical materials realizes
\hbox{$S\!=\!1/2$} two-leg ladder systems with different leg interactions.

In the following discussions, we confine ourselves to the case
where $J_{\rm r}$ is ferromagnetic, and we put \hbox{$J_{\rm r}\!=\!-1$},
choosing $|J_{\rm r}|$ as the unit of energy.  Then, when
\hbox{$0\!<\!|J_{{\rm l},\ell}|\!\ll\!1$}, the present ladder system can be
mapped onto the \hbox{$S\!=\!1$} chain by using the degenerate perturbation
theory.  We discuss this mapping in the next section (section~2).  Section~3
is devoted to the discussions on the ground-state phase diagram.  Assuming,
for simplicity, that \hbox{$J_{{\rm l},a}\!=-0.2$} or $0.2$ and
\hbox{$0\!\leq\!\Delta\!<\!1$} (the $XY$-type anisotropy of rung
interactions), we determine the ground-state phase diagrams on the $\Delta$
versus $J_{{\rm l},b}$ plane.  We mainly use the numerical methods such as the
exact-diagonalization (ED) method and the density-matrix renormalization-group
(DMRG) method~\cite{dmrg-white-1,dmrg-white-2} with the help of physical
considerations.  Finally, we give concluding remarks in section~4. 

\section{Mapping onto the \mbox{\boldmath{$S\!=\!1$}} chain}

We discuss the case where \hbox{$0\!<\!|J_{{\rm l},\ell}|\!\ll\!|J_{\rm r}|$},
assuming that \hbox{$J_{\rm r}\!=\!-1$}.  The four eigenstates for rung $j$
are given by  \hbox{$\psi_j^{(1,+)}\!=\!\alpha_{j,a}\alpha_{j,b}$},
\hbox{$\psi_j^{(1,0)}\!=\!(\alpha_{j,a}\beta_{j,b}\!+\!\beta_{j,a}\alpha_{j,b})
/{\sqrt{2}}$},
\hbox{$\psi_j^{(1,-)}\!=\!\beta_{j,a}\beta_{j,b}$} and
\hbox{$\psi_j^{(0,0)}\!=\!(\alpha_{j,a}\beta_{j,b}\!-\!\beta_{j,a}\alpha_{j,b})
/{\sqrt{2}}$}, and the corresponding energies are, respectively,
$E^{(1,+)}\!=\!-\Delta/4$, $E^{(1,0)}\!=\!(\Delta\!-\!2)/4$,
$E^{(1,-)}\!=\!-\Delta/4$ and $E^{(0,0)}\!=\!(\Delta\!+\!2)/4$, for all
$j$'s.  Thus, the state $\psi_j^{(0,0)}$ can be neglected.  We
introduce the pseudo \hbox{$S\!=\!1$} operator ${\vec T}_j$ for rung $j$, and
make the $T_j^z\!=\!+1$, $0$ and $-1$ states correspond to the
$\psi_j^{(1,+)}$, $\psi_j^{(1,0)}$ and $\psi_j^{(1,-)}$ states, respectively.
The relation ${\vec T}_j\!=\!{\vec S}_{j,a}\!+\!{\vec S}_{j,b}$ holds, as is
readily shown by comparing the matrix elements of both operators
${\vec T}_j$ and ${\vec S}_{j,l}$ in the subspace of
$\phi_j^{(1,+)}$, $\phi_j^{(1,0)}$ and $\phi_j^{(1,-)}$.  Thus, the
Hamiltonian~(\ref{eq:hamiltonian}) for the \hbox{$S\!=\!1/2$} operator
${\vec S}_{j,l}$ can be mapped onto the effective Hamiltonian
${\cal H}_{{\rm eff}}$ for the \hbox{$S\!=\!1$} operator ${\vec T}_j$, which
is given by
\begin{equation}
    {\cal H}_{{\rm eff}}
         = J_{\rm eff} \sum_{j=1}^L {\vec T}_{j} \cdot{\vec T}_{j+1}
                   + D_{\rm eff} \sum_{j=1}^L (T_j^z)^2 \,;\quad
       J_{\rm eff} = \frac{J_{{\rm l},a} + J_{{\rm l},b}}{4}\,,\quad
       D_{\rm eff} = \frac{(1-\Delta)}{2},
\label{eq:effective}
\end{equation}
where $T_j^z$ is the $z$-component of ${\vec T}_j$.  It is noted that the
on-site anisotropy ($D_{\rm eff}$-) term comes from the difference between
$E^{(1,+)}=E^{(1,-)}$ and $E^{(1,0)}$.

The above ${\cal H}_{{\rm eff}}$ is the result of the degenerate perturbation
calculation in the lowest-order of $|J_{{\rm l},\ell}|/|J_{\rm r}|$.  It is
apparent that this is not applicable to discussing the frustrated region of
the original Hamiltonian~(\ref{eq:hamiltonian}), which includes the case of
\hbox{$J_{{\rm l},a}\!+\!J_{{\rm l},b}\!=\!0$}.  In order to improve this
point, higher-order perturbation calculations are indispensable; these
calculations are left for a future study.

The ground-state phase diagram of the anisotropic \hbox{$S\!=\!1$} chain has
been determined by several authors~\cite{chen-etal,hu-etal,wierschem-etal}.
According to their results, as the value of $D_{\rm eff}$ increases
from zero, the phase transition from
the $XY$ (or Haldane) phase to the large-$D$ phase takes place at
\hbox{$D_{\rm eff}\!\simeq\!1$} when \hbox{$J_{\rm eff}\!=-1$} (or when
\hbox{$J_{\rm eff}\!=1$}).  Thus, we may expect that in our \hbox{$S\!=\!1/2$}
ladder with \hbox{$J_{\rm r}\!=\!-1$}, the phase transition between the
$XY$ and TD phases occurs at
\hbox{$J_{{\rm l},a}\!+\!J_{{\rm l},b}\!\simeq\!2(\Delta\!-\!1)$} when
\hbox{$J_{{\rm l},a}\!+\!J_{{\rm l},b}\!<\!0$} (or, equivalently, when
\hbox{$\Delta\!<\!1$}), and also that the phase transition between the
Haldane and TD phases occurs at
\hbox{$J_{{\rm l},a}\!+\!J_{{\rm l},b}\!\simeq\!2(1\!-\!\Delta\!)$} when
\hbox{$J_{{\rm l},a}\!+\!J_{{\rm l},b}\!>\!0$} (or, when
\hbox{$\Delta\!<\!1$}, again).  It is noted that the large-$D$ state in the
spin-1 chain is equivalent to the TD state in the present \hbox{$S\!=\!1/2$}
ladder, since in the valence bond picture of the former state, each
\hbox{$S\!=\!1$} spin consists of two \hbox{$S\!=\!1/2$} spins forming the TD
pair, as is well known.

\section{Ground-state phase diagrams}

\begin{figure}[t]
  \scalebox{0.28}{\includegraphics{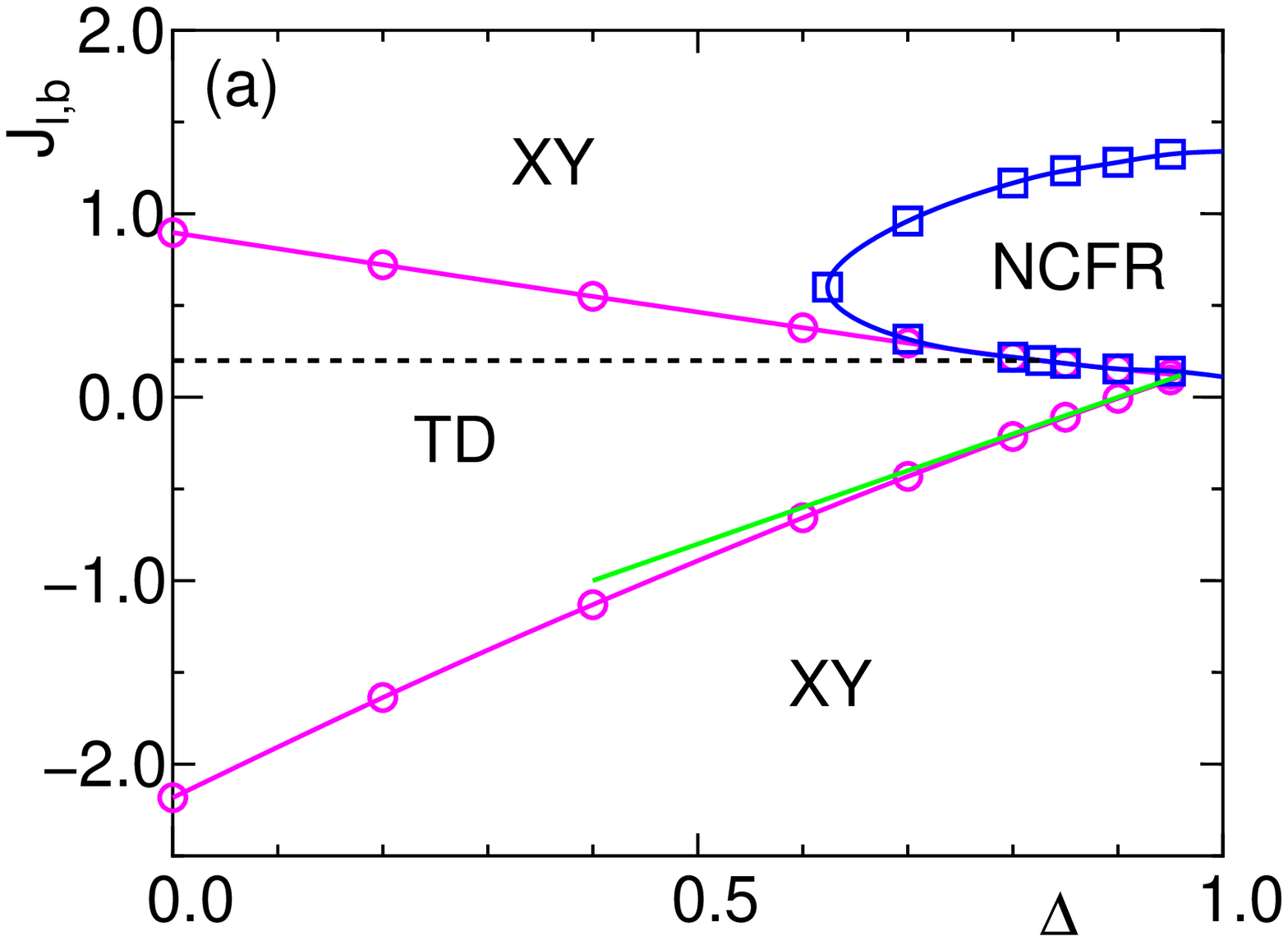}}~~
  \scalebox{0.28}{\includegraphics{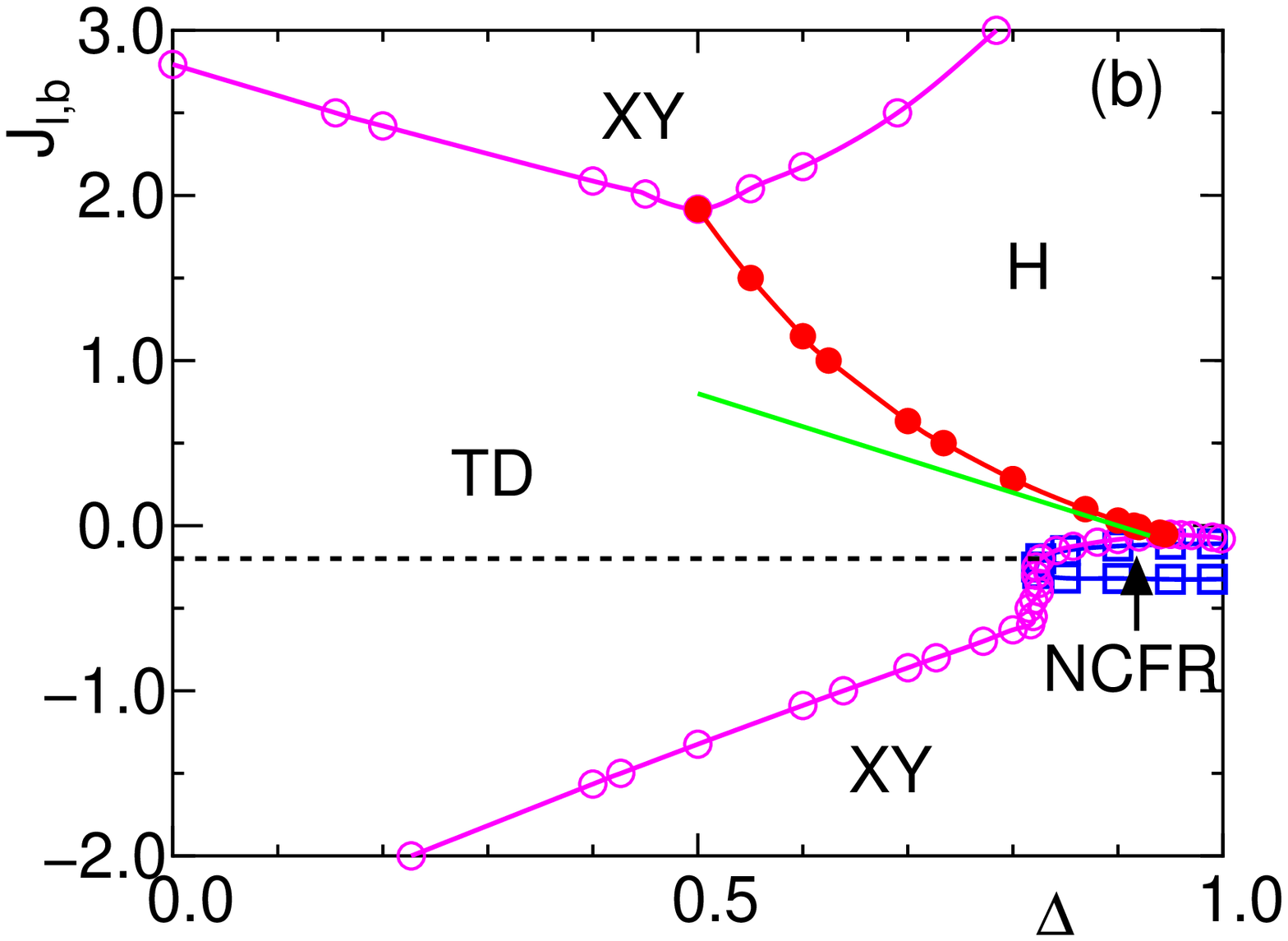}}~~
  \scalebox{0.28}{\includegraphics{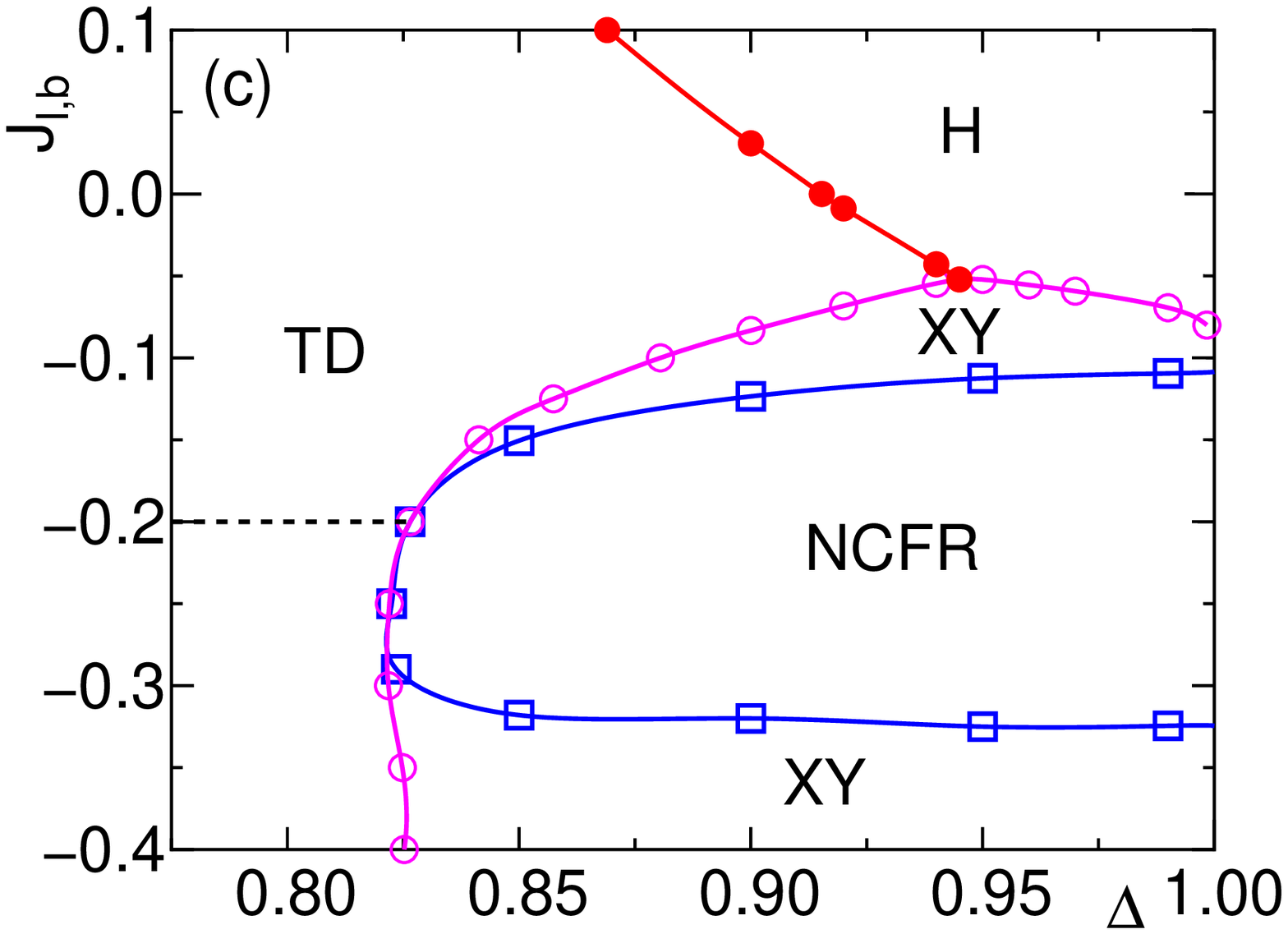}}
  \caption{Ground-state phase diagram on the $\Delta$ versus $J_{{\rm l},b}$
           plane for (a) \hbox{$J_{{\rm l},a}\!=\!-0.2$} and (b,c)
           \hbox{$J_{{\rm l},a}\!=\!0.2$} with \hbox{$J_{{\rm r}}\!=\!-1$}; in
           (c) part of (b) is enlarged.  The regions designated by TD, XY, NCFR
           and H are, respectively, those of the triplet-dimer, $XY$,
           non-collinear ferrimagnetic and Haldane phases.  See the text for
           the meanings of several lines.
           }
\label{fig:phase-diagram}
\end{figure}

Throughout this section we assume that \hbox{$J_{{\rm r}}\!=\!-1$}, as
mentioned before. Figure~\ref{fig:phase-diagram} shows the ground-state phase
diagrams on the $\Delta$ versus $J_{{\rm l},b}$ plane determined for
\hbox{$J_{{\rm l},a}\!=\!-0.2$} and \hbox{$J_{{\rm l}.a}\!=\!0.2$}.
The former phase diagram consists of the TD, $XY$ and non-collinear
ferrimagnetic (NCFR) phases~\cite{tsukano-takahashi,yoshikawa-miyashita}, and
in the latter one, the Haldane (H) phase appears in addition to the above
three phases.  There are three kinds of the phase transition lines, which we
have numerically estimated as discussed below in detail.  The magenta lines
with open circles are the phase transition lines between the TD or H phase and
the $XY$ phase which are of the
Berezinskii-Kosterlitz-Thouless (BKT) type~\cite{BKT-1,BKT-2}, the red line
with closed circles is the phase transition line between the TD and H phases
which are of the Gaussian-type, and finally the blue lines with open squares
are the phase
transition lines between the NCFR phase and the TD or $XY$ phase.  In the
latter phase diagram, there are two tricritical points at
\hbox{$(\Delta, J_{{\rm l},b})\!=\!\bigl(0.500(1),1.917(1)\bigr)$}
and $\bigl(0.945(1),-0.052(1)\bigr)$ associated with TD, $XY$ and H phases. The
green straight lines show the results of the comparison of the degenerate
perturbation calculations with the numerical
results~\cite{chen-etal,hu-etal,wierschem-etal} (see section~2); in
Fig.~\ref{fig:phase-diagram}(a) it is for the TD-$XY$ transition and given
by \hbox{$J_{{\rm l},b}\!=\!2\Delta\!-\!1.8$}, while in
Fig.~\ref{fig:phase-diagram}(b) it is for the TD-H transition and given
by \hbox{$J_{{\rm l},b}\!=\!-2\Delta\!+\!1.8$}.  In both cases they are in
excellent agreement with the numerical results at least when $|J_{{\rm l},b}|$
is not too large.  It is noted that on the special lines where 
\hbox{$J_{{\rm l},a}\!+\!J_{{\rm l},b}\!=\!0$}, which are shown by the black
broken lines, the direct-product TD state is the {\it exact ground state}.

In the following explanations for the estimation of the above phase boundary
lines, we denote, respectively, by $E_0(L,M;{\rm pbc})$ and
$E_1(L,M;{\rm pbc})$ the lowest and second-lowest energy eigenvalues of the
Hamiltonian~(\ref{eq:hamiltonian}) within the subspace determined by $L$ and
$M$ under periodic boundary conditions,
\hbox{${\vec S}_{L+1,\ell}\!=\!{\vec S}_{1,\ell}$}.  The quantity $M$ is the
total magnetization given by
\hbox{$M\!=\!\sum_{j=1}^{L} (S_{j,a}^z\!+\!S_{j,b}^z)$}, which is a good
quantum number with the eigenvalues of \hbox{$M\!=\!0$}, $\pm 1$,
$\cdots$, $\pm L$.  Similarly, we also denote by $E_0(L,M,P;{\rm tbc})$ the
lowest energy eigenvalue of the Hamiltonian~(\ref{eq:hamiltonian}) within the
subspace determined by $L$, $M$ and $P$ under twisted boundary conditions,
\hbox{$S_{L+1,\ell}^x\!=\!-S_{1,\ell}^x$},
\hbox{$S_{L+1,\ell}^y\!=\!-S_{1,\ell}^y$} and
\hbox{$S_{L+1,\ell}^z\!=\!S_{1,\ell}^z$}, where \hbox{$P(=\!+1$} or $-1)$ is
the eigenvalue of the space inversion operator with respect to the twisted
bond, \hbox{${\vec S}_{j,\ell}\!\leftrightarrow\!{\vec S}_{L+1-j,\ell}$}.  We
further denote by $E_0(L,M;{\rm obc})$ the lowest energy eigenvalue of the
Hamiltonian~(\ref{eq:hamiltonian}) within the subspace determined by $L$ and
$M$ under open boundary conditions, where the sums over $j$ for leg
interactions  are taken from \hbox{$j\!=\!1$} to \hbox{$L\!-\!1$}.

The most powerful method to estimate numerically the phase boundary lines
between two of the TD, $XY$ and H phases is the level spectroscopy (LS)
method developed by Okamoto, Nomura and Kitazawa~\cite{LSmethod-1,LSmethod-2,
LSmethod-3, LSmethod-4}.  In this method, the following three excitation
energies~\cite{comment-1},
\hbox{$\Delta E_{02}^{({\rm p})}(L)\!=\!E_0(L,2;{\rm pbc})\!
-\!E_0(L,0;{\rm pbc})$},
\hbox{$\Delta E_{00}^{({\rm p},{\rm t})}(L,+1)\!=\!E_0(L,0,+1;{\rm tbc})\!
-\!E_0(L,0;{\rm pbc})$} and
\hbox{$\Delta E_{00}^{({\rm p},{\rm t})}(L,-1)\!=\!E_0(L,0,-1;{\rm tbc})\!
-\!E_0(L,0;{\rm pbc})$} should be compared in the thermodynamic
(\hbox{$L\!\to\!\infty$)} limit.  More strictly speaking, the critical value
$J_{{\rm l},b\,({\rm cr})}^{(XY,{\rm TD})}$ of the BKT $XY$-TD transition,
the critical value $J_{{\rm l},b\,({\rm cr})}^{(XY,{\rm H})}$ of the BKT
$XY$-H transition and the critical value
$J_{{\rm l},b\,({\rm cr})}^{({\rm TD,H})}$ of the Gaussian TD-H transition,
which are all for given values of $J_{{\rm l},a}$ and $\Delta$, are estimated
as follows.  First, the corresponding finite-size critical values
$J_{{\rm l},b\,({\rm cr})}^{(XY,{\rm TD})}(L)$,
$J_{{\rm l},b\,({\rm cr})}^{(XY,{\rm H})}(L)$ and
$J_{{\rm l},b\,({\rm cr})}^{({\rm TD,H})}(L)$ are estimated, respectively,
by solving numerically the equations~\cite{comment-3},
\begin{eqnarray}
   && \Delta E_{02}^{({\rm p})}(L) = \Delta E_{00}^{({\rm p},{\rm t})}(L,+1)
                         < \Delta E_{00}^{({\rm p},{\rm t})}(L,-1)\,,    \\
   && \Delta E_{02}^{({\rm p})}(L) = \Delta E_{00}^{({\rm p},{\rm t})}(L,-1)
                         < \Delta E_{00}^{({\rm p},{\rm t})}(L,+1)\,,    \\
   && \Delta E_{00}^{({\rm p},{\rm t})}(L,+1)
                         = \Delta E_{00}^{({\rm p},{\rm t})}(L,-1)
                         < \Delta E_{02}^{({\rm p})}(L)\,.
\end{eqnarray}
Then, these finite-size results are extrapolated to the
\hbox{$L\!\to\!\infty$} limit to obtain, respectively, the critical
values, $J_{{\rm l},b\,({\rm cr})}^{(XY,{\rm TD})}$, 
$J_{{\rm l},b\,({\rm cr})}^{(XY,{\rm H})}$ and
$J_{{\rm l},b\,({\rm cr})}^{({\rm TD,H})}$.

\begin{figure}[b]
  \scalebox{0.28}{\includegraphics{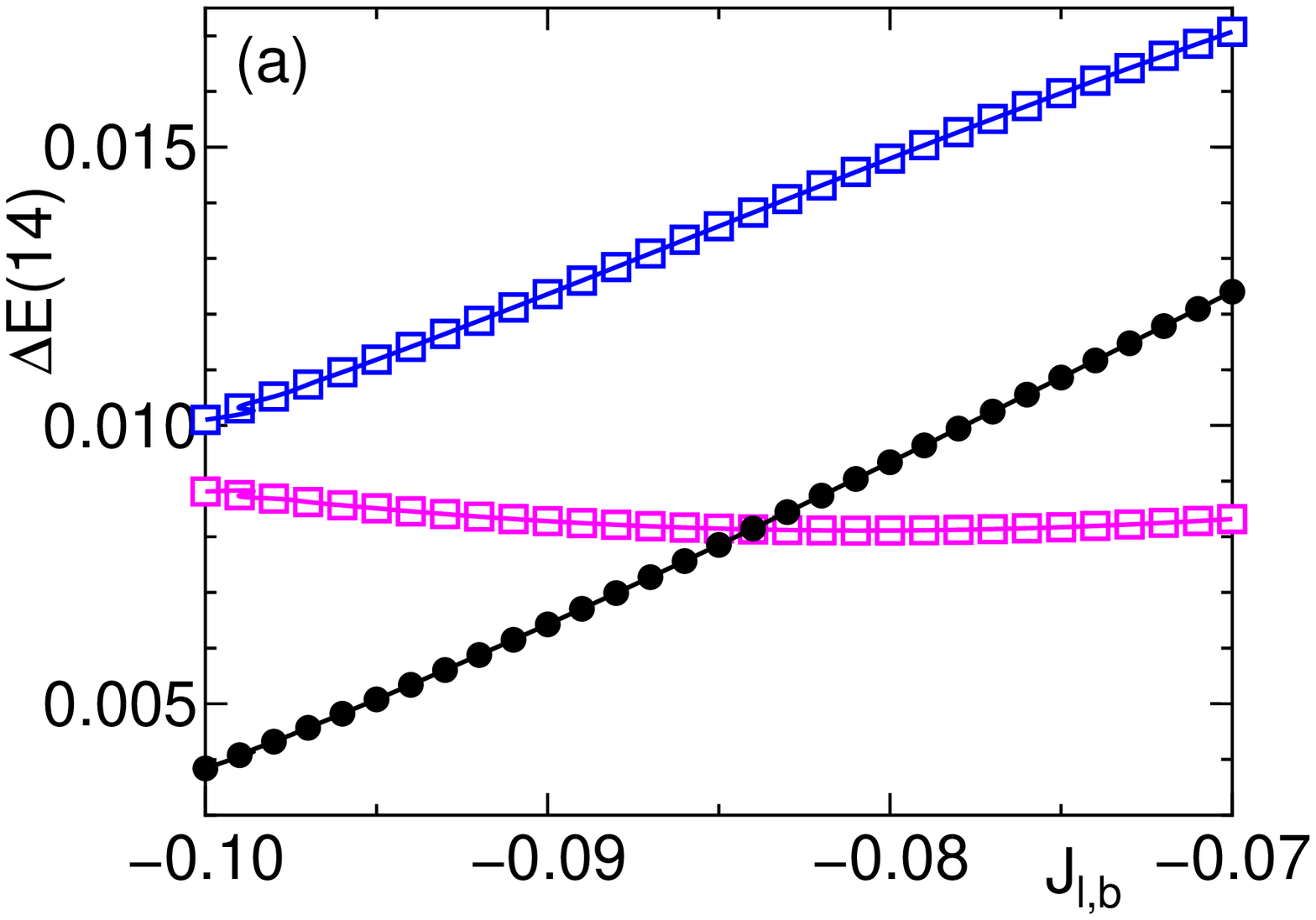}}~~
  \scalebox{0.28}{\includegraphics{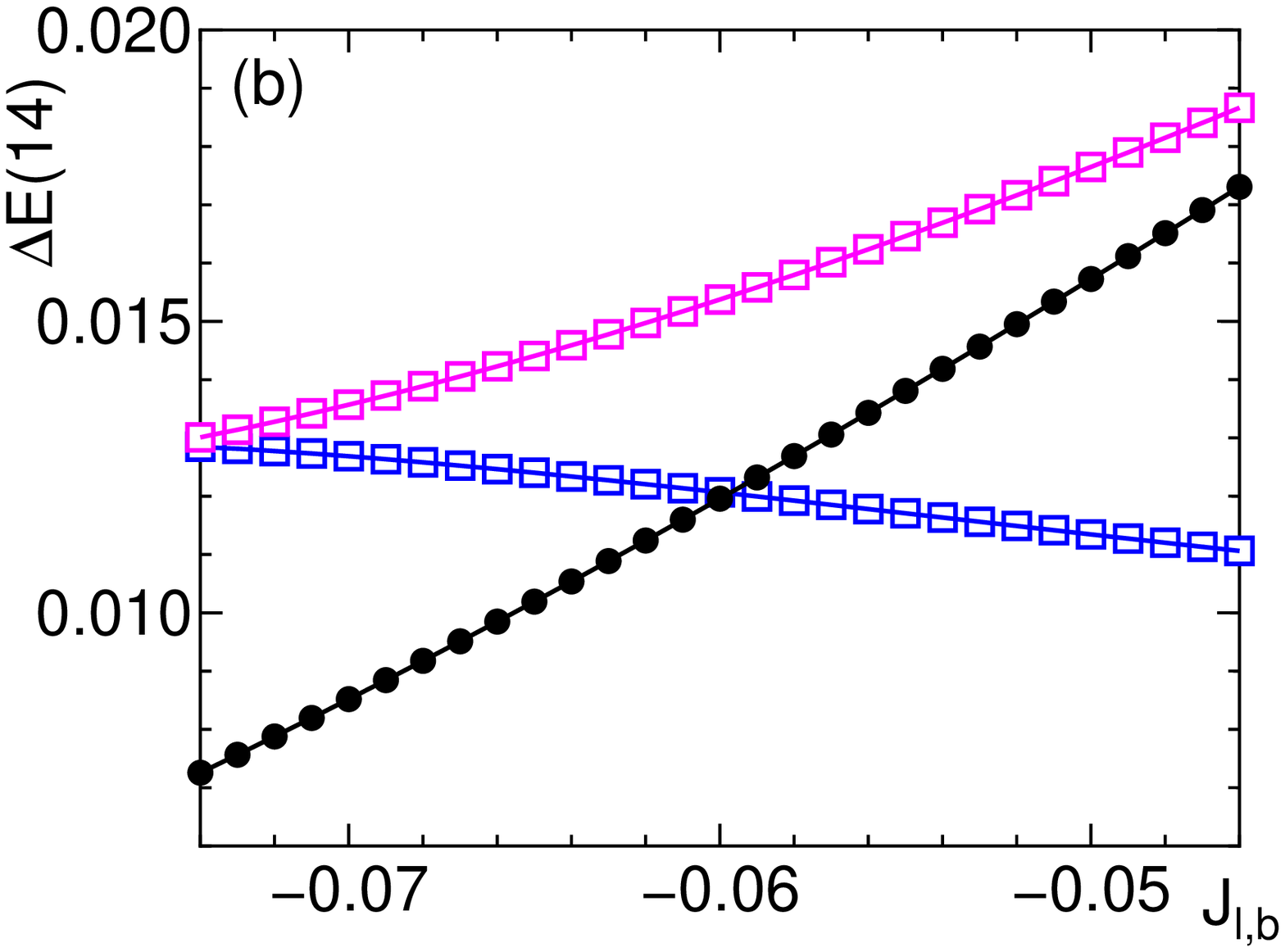}}~~
  \scalebox{0.28}{\includegraphics{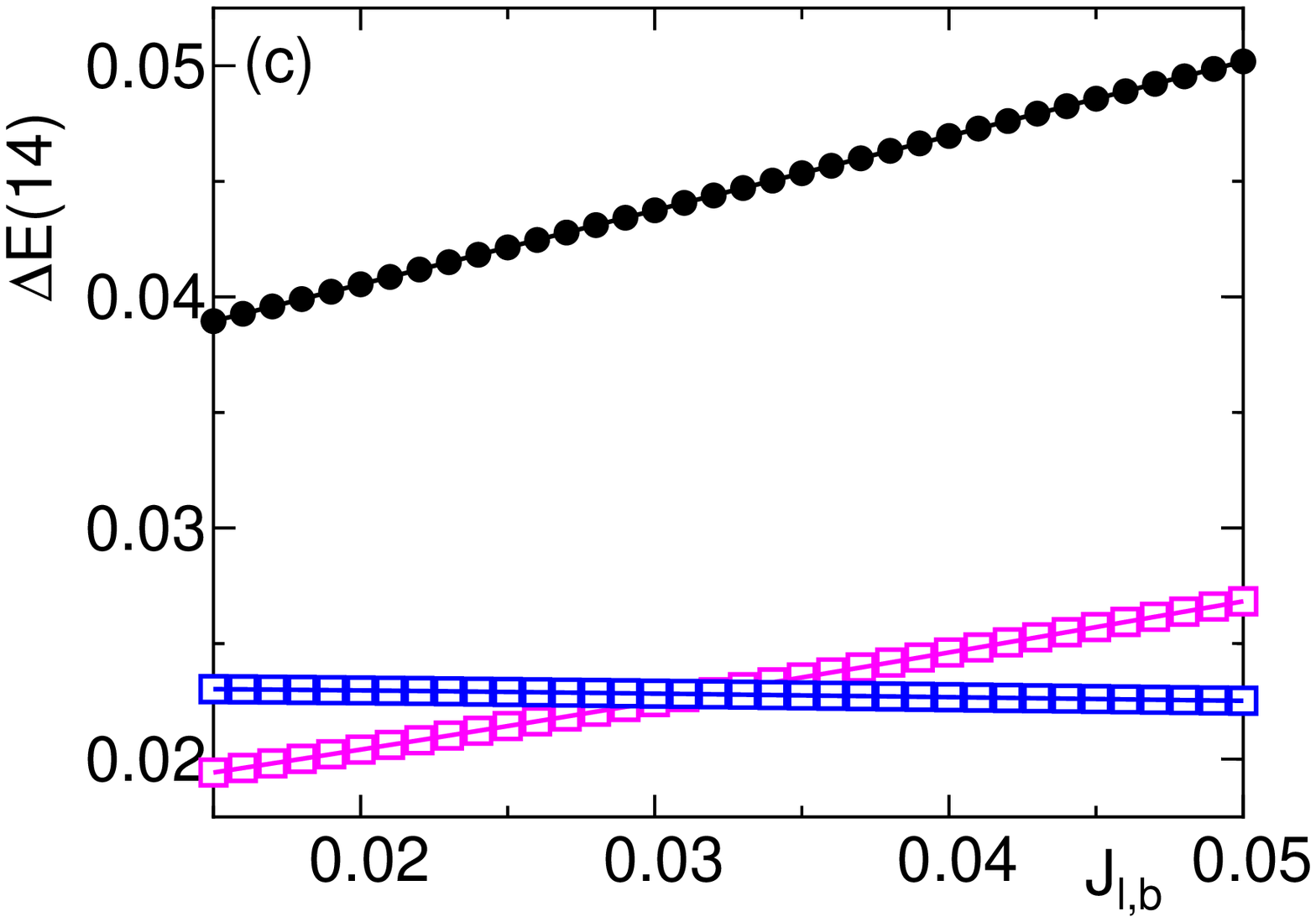}}
  \caption{Examples of the $\Delta E_{02}^{({\rm p})}(14)$ (black closed
           circles), $\Delta E_{00}^{({\rm p},{\rm t})}(14,+1)$ (magenta open
           squares) and  $\Delta E_{00}^{({\rm p},{\rm t})}(14,-1)$ (blue open
           squares) versus
           $J_{{\rm l},b}$ curves for (a,c) \hbox{$\Delta\!=\!0.90$}
           and (b) \hbox{$\Delta\!=\!0.97$} with
           \hbox{$J_{{\rm l},a}\!=\!0.2$} and \hbox{$J_{\rm r}\!=\!-1$}
           $\bigl($see Fig.~$\!$\ref{fig:phase-diagram}(c).$\bigr)$  In
           (a), (b) and (c) we obtain, respectively,
           \hbox{$J_{{\rm l},b\,({\rm cr})}^{(XY,{\rm TD})}(14)\!=\!-0.084049$}
           from the crossing point of the black and magenta curves, 
           \hbox{$J_{{\rm l},b\,({\rm cr})}^{(XY,{\rm H})}(14)\!=\!-0.059748$}
           from the crossing point of the black and blue curves and 
           \hbox{$J_{{\rm l},b\,({\rm cr})}^{({\rm TD,H})}(14)\!=\!0.031604$}
           from the crossing point of the magenta and blue curves.
           }
\label{fig:excitation-energy}
\end{figure}
\begin{figure}[t]
  \vspace{0.4cm}
  \scalebox{0.26}{\includegraphics{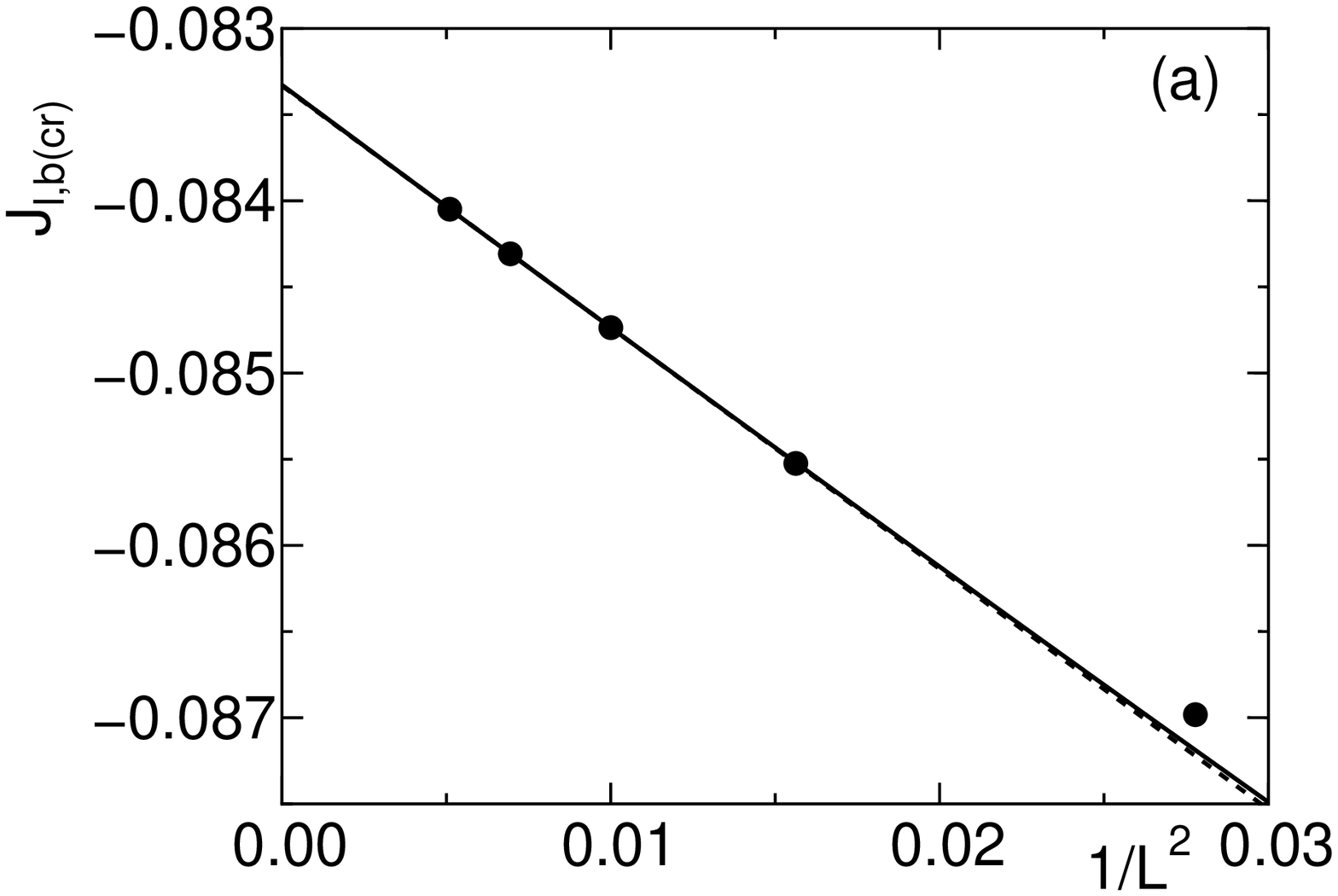}}~~
  \scalebox{0.26}{\includegraphics{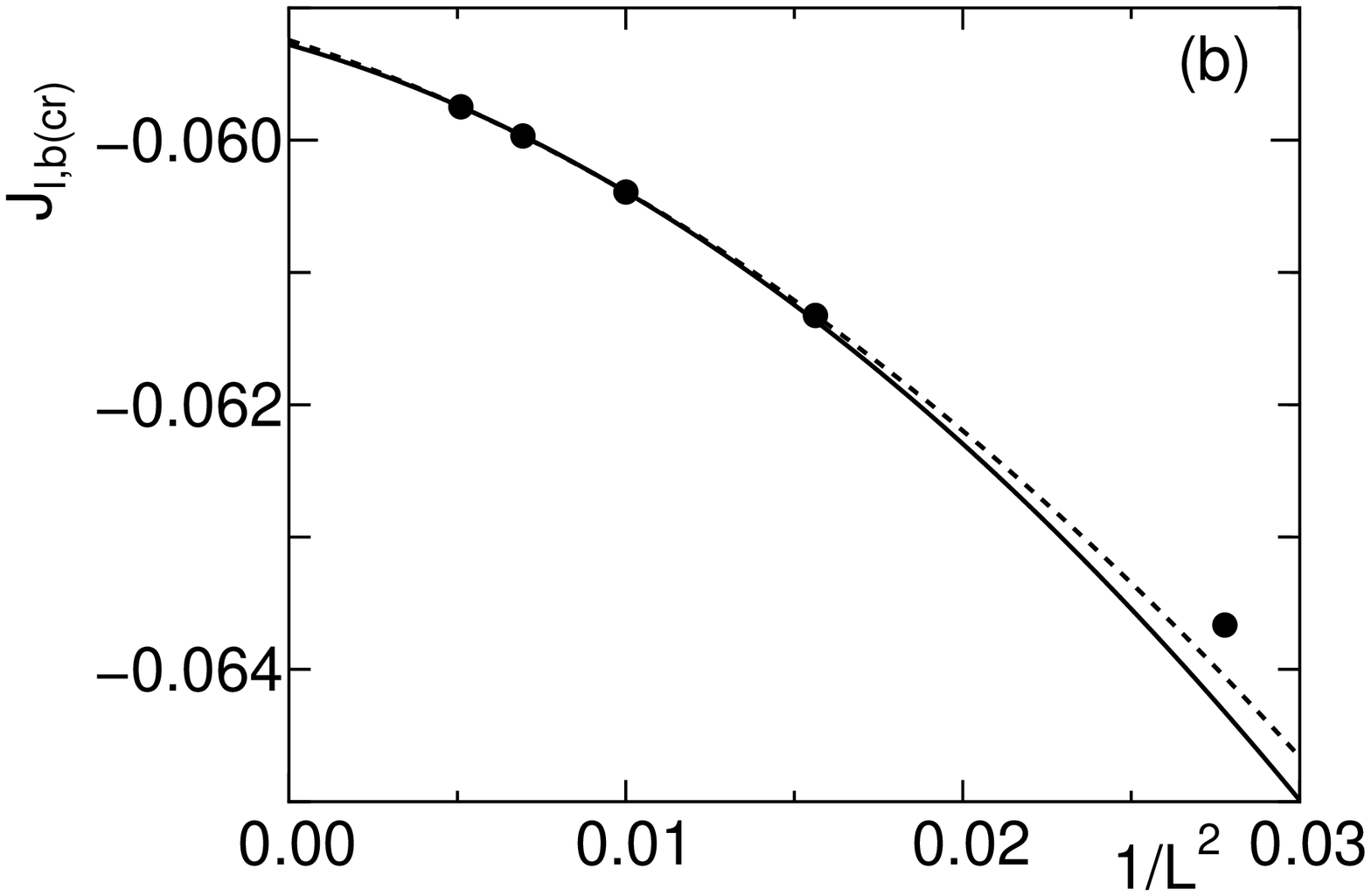}}~~
  \scalebox{0.26}{\includegraphics{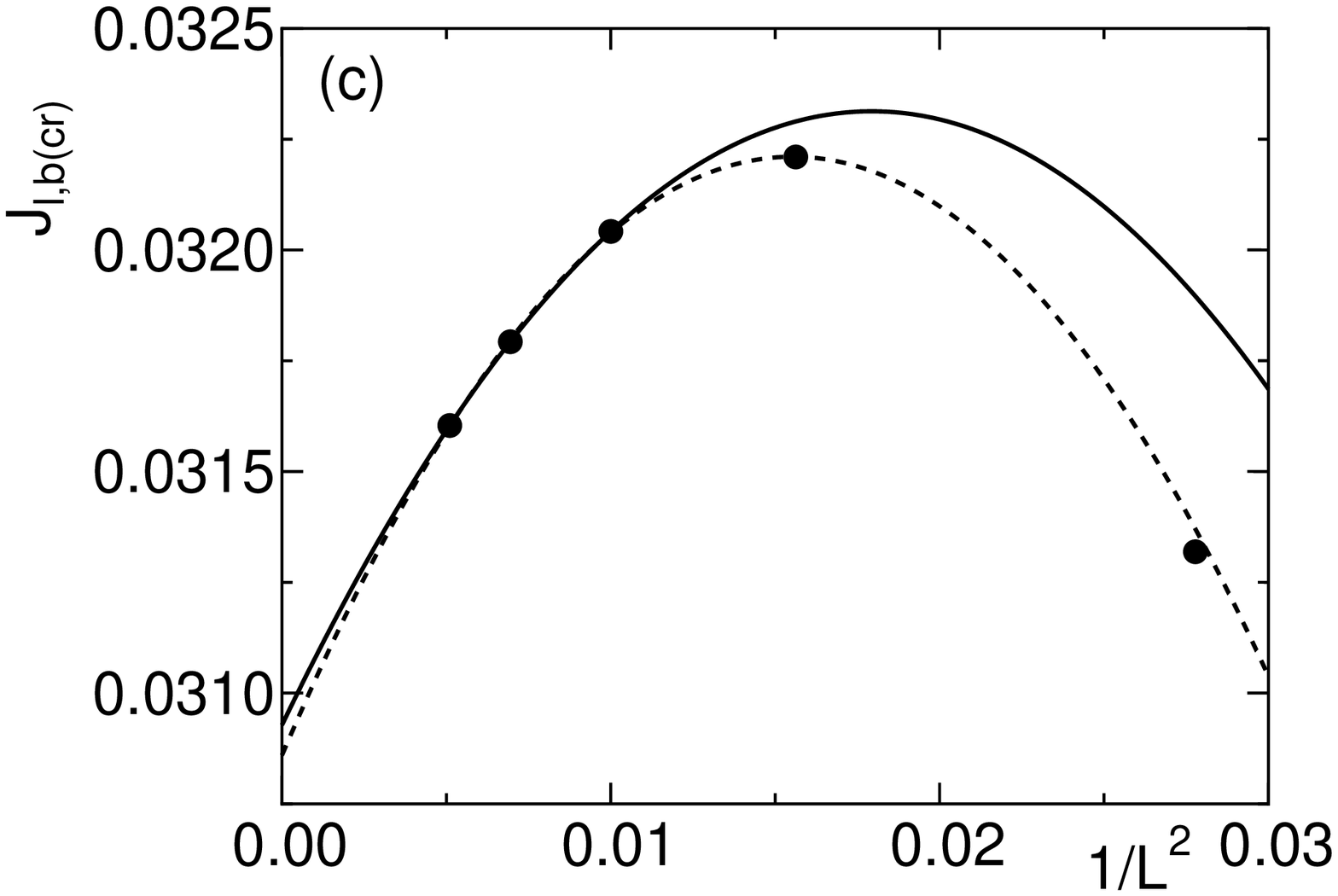}}
  \caption{Examples of the \hbox{$L\!\to\!\infty$} extrapolations of (a)
          $J_{{\rm l},b\,({\rm cr})}^{(XY,{\rm TD})}(L)$ for
          \hbox{$\Delta\!=\!0.90$}, (b) 
          $J_{{\rm l},b\,({\rm cr})}^{(XY,{\rm H})}(L)$ for
          \hbox{$\Delta\!=\!0.97$} and (c)
          $J_{{\rm l},b\,({\rm cr})}^{({\rm TD,H})}(L)$ for
          \hbox{$\Delta\!=\!0.90$}, where
          \hbox{$J_{{\rm l},a}\!=\!0.2$} and \hbox{$J_{\rm r}\!=\!-1$}
          $\bigl($see Fig.~\ref{fig:phase-diagram}(c).$\big)$  We assume that
          these finite-size critical values are quadratic functions of $1/L^2$.
          The broken lines represent the least-square fittings by use of
          \hbox{$L\!=\!14$}, $12$, $10$ and $8$ data, while the solid lines
          those without \hbox{$L\!=\!8$} data.  From these extrapolations,
          we obtain
          \hbox{$J_{{\rm l},b\,({\rm cr})}^{(XY,{\rm TD})}\!=\!-0.0833(1)$}
          in (a),
          \hbox{$J_{{\rm l},b\,({\rm cr})}^{(XY,{\rm H})}\!=\!-0.0593(1)$}
          in (b) and
          \hbox{$J_{{\rm l},b\,({\rm cr})}^{({\rm TD,H})}\!=\!0.0309(1)$}
          in (c), where the numerical errors are estimated from the
          difference between the extrapolated results with and without the
          \hbox{$L\!=\!8$} data.
          }
\label{fig:extrapolation}
\end{figure}

Practically, we have made the ED calculations to estimate
$J_{{\rm l},b\,({\rm cr})}^{(XY,{\rm TD})}(L)$,
$J_{{\rm l},b\,({\rm cr})}^{(XY,{\rm H})}(L)$ and
$J_{{\rm l},b\,({\rm cr})}^{({\rm TD,H})}(L)$ for finite-$L$ systems with
\hbox{$2L\!=\!12$}, $16$, $\cdots$, $28$ spins.  The procedures for these
estimations are shown in Fig.~\ref{fig:excitation-energy}, for example, for
\hbox{$L\!=\!14$}, \hbox{$J_{{\rm l},a}\!=\!0.2$} and \hbox{$\Delta\!=\!0.90$}
or $0.97$.  Performing the \hbox{$L\!\to\!\infty$} extrapolations of the above
finite-size critical values, we have fitted them to quadratic functions of
$1/L^2$ by use of the least-square method, as explained in
Fig.~\ref{fig:extrapolation}, for example, for 
\hbox{$J_{{\rm l},a}\!=\!0.2$} and \hbox{$\Delta\!=\!0.90$} or $0.97$, again.
Then, as the results of the extrapolations, we have obtained in the
\hbox{$J_{{\rm l},a}\!=\!0.2$} case,
\hbox{$J_{{\rm l},b\,({\rm cr})}^{(XY,{\rm TD})}\!=\!-0.0833(1)$} and
\hbox{$J_{{\rm l},b\,({\rm cr})}^{({\rm TD,H})}\!=\!0.0309(1)$} for
\hbox{$\Delta\!=\!0.90$}, and also
\hbox{$J_{{\rm l},b\,({\rm cr})}^{(XY,{\rm H})}\!=\!-0.0593(1)$}
for \hbox{$\Delta\!=\!0.97$}.  The phase transition lines shown by the magenta
and red lines in Fig.~\ref{fig:phase-diagram}(b) and (c) are drawn by plotting,
as functions of $\Delta$, the values of
$J_{{\rm l},b\,({\rm cr})}^{(XY,{\rm TD})}$,
$J_{{\rm l},b\,({\rm cr})}^{({\rm TD,H})}$, and
$J_{{\rm l},b\,({\rm cr})}^{(XY,{\rm H})}$ calculated for various
values of $\Delta$.  Similarly, the phase transition lines shown by the magenta
lines in Fig.~\ref{fig:phase-diagram}(a) are obtained by calculating
$J_{{\rm l},b\,({\rm cr})}^{(XY,{\rm TD})}$ and
$J_{{\rm l},b\,({\rm cr})}^{({\rm TD,H})}$ for various values of $\Delta$
in the case of \hbox{$J_{{\rm l},a}\!=\!-0.2$}.

\begin{figure}[b]
  \vspace{0.5cm}
  \scalebox{0.28}{\includegraphics{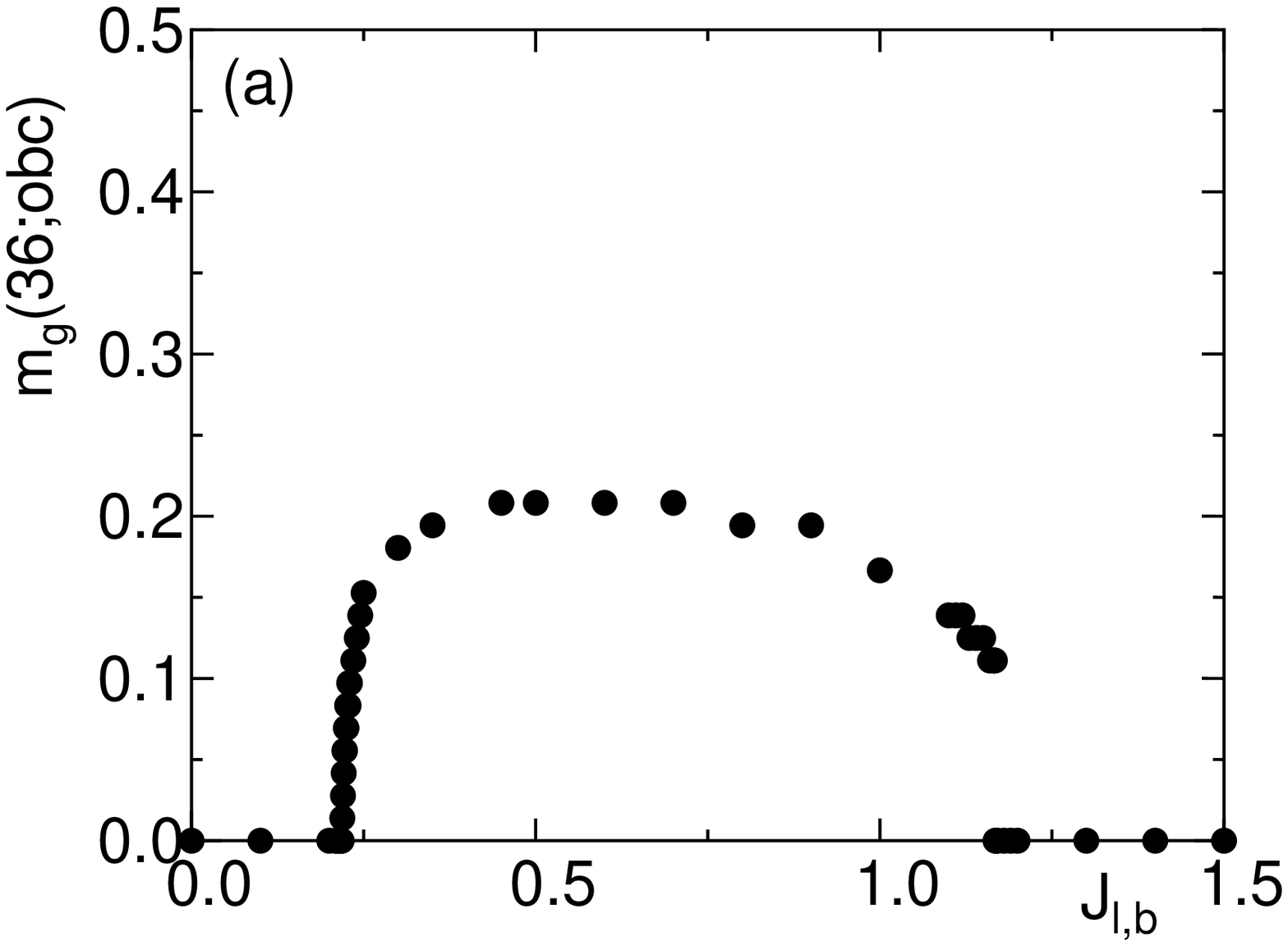}}~~
  \scalebox{0.28}{\includegraphics{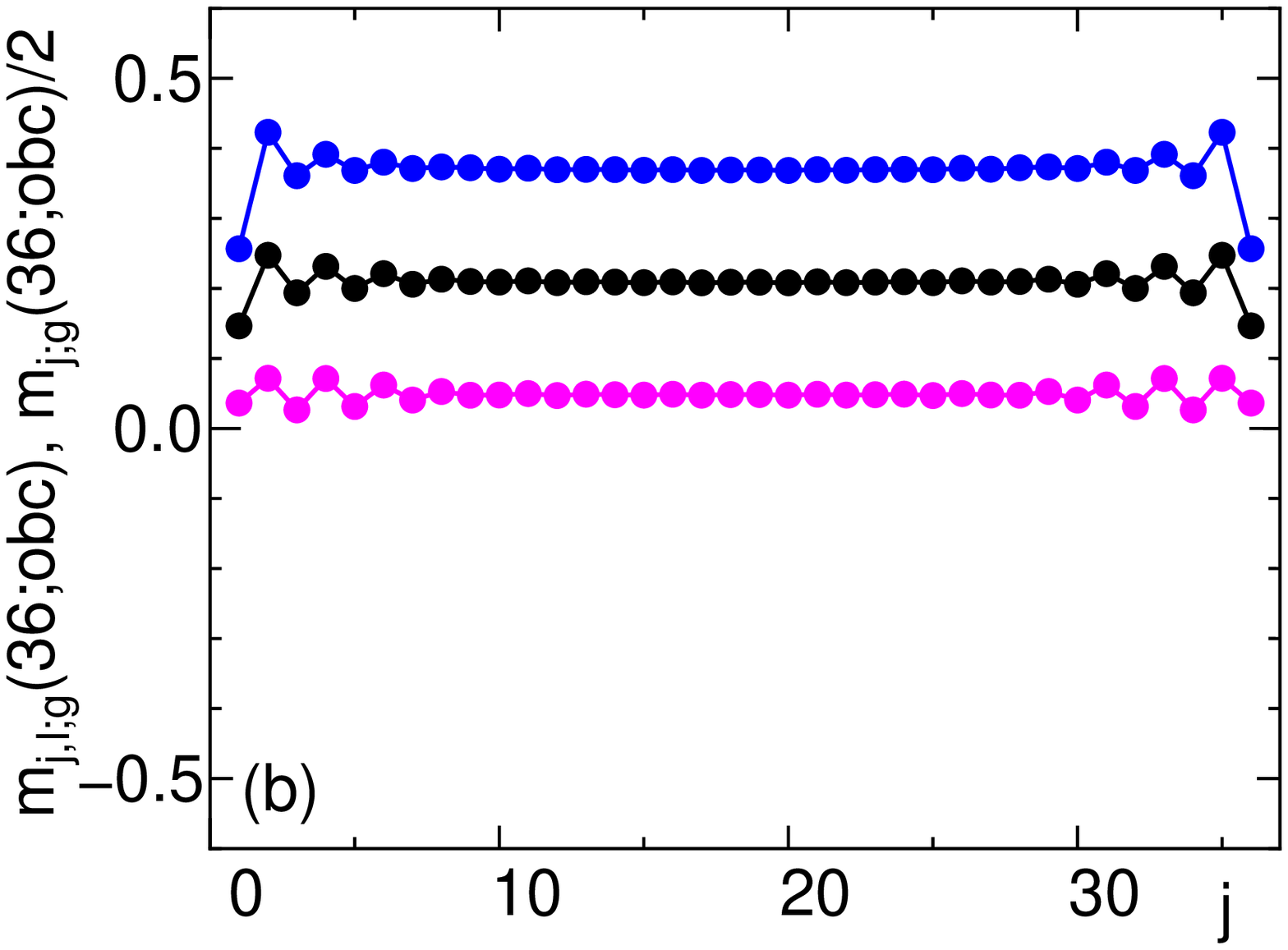}}~~
  \scalebox{0.28}{\includegraphics{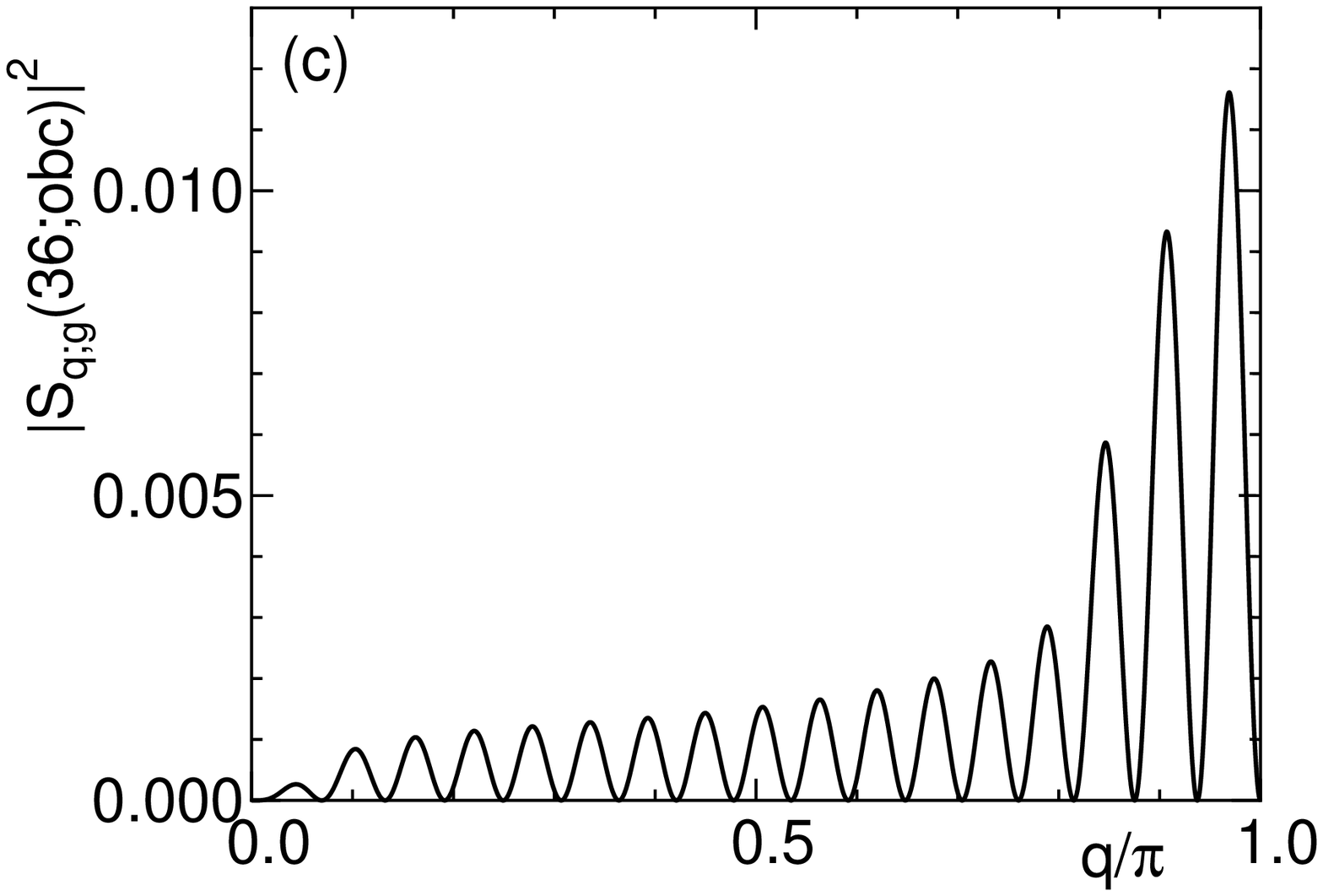}}
  \caption{(a)~Plot of $m_{\rm g}(36;{\rm obc})$ versus $J_{{\rm l},b}$,
          (b)~those of $m_{j;{\rm g}}(36;{\rm obc})/2$ (black line),
          $m_{j,a;{\rm g}}(36;{\rm obc})$ (blue line) and
          $m_{j,b;{\rm g}}(36;{\rm obc})$ (magenta line) versus $j$ and
          (c)~that of $|S_{q;{\rm g}}(36;{\rm obc})|^2$ versus $q/\pi$.
          The quantity $m_{\rm g}(36;{\rm obc})$ in (a) is obtained in the
          case where \hbox{$J_{{\rm l},a}\!=\!-0.2$},
          \hbox{$\Delta\!=\!0.8$} and \hbox{$J_{\rm r}\!=\!-1$}, while
          those $m_{j;{\rm g}}(36;{\rm obc})$ and
          $|S_{q;{\rm g}}(36;{\rm obc})|^2$ in (b,c) are obtained in the case
          where \hbox{$J_{{\rm l},a}\!=\!-0.2$},
          \hbox{$J_{{\rm l},b}\!=\!0.7$}, \hbox{$\Delta\!=\!0.8$} and
          \hbox{$J_{\rm r}\!=\!-1$}.
          $\bigl($see Fig.~\ref{fig:phase-diagram}(a).$\big)$
          }
\label{fig:GS-magnetization-etc}
\end{figure}

Let us denote by $M_{\rm g}(L;{\rm obc})$ the ground-state magnetization for
the system with $2L$ spins under open boundary conditions, which is
the value of $M$ giving the lowest value of $E_0(L,M;{\rm obc})$'s.  In the
NCFR phase, $M_{\rm g}(L;{\rm obc})$ is
finite $\bigl($\hbox{$0\!<\!M_{\rm g}(L;{\rm obc})\!<\!L$}$\bigr)$, while in
other phases, \hbox{$M_{\rm g}(L;{\rm obc})\!=\!0$}~\cite{comment-1}.  We
have carried out DMRG calculations~\cite{dmrg-white-1,dmrg-white-2} for the
finite system with \hbox{$2L\!=\!72$} spins to estimate the ground-state
magnetization per spin, $m_{\rm g}(L;{\rm obc})$, which is defined by
\hbox{$m_{\rm g}(L;{\rm obc})\!=\!M_{\rm g}(L;{\rm obc})/(2L)$}.  The obtained
results in the case where
\hbox{$J_{{\rm l},a}\!=\!-0.2$} and \hbox{$\Delta\!=\!0.8$} are depicted in
Fig.~\ref{fig:GS-magnetization-etc}(a).  We see from this figure that the phase
transition from the TD phase to the NCFR phase and that from the NCFR
phase to the $XY$ phase successively occur with increasing $J_{{\rm l},b}$
$\bigl($see Fig.~\ref{fig:phase-diagram}(a)$\bigr)$.  The finite-size critical
values for the former and latter transitions in the
\hbox{$J_{{\rm l},a}\!=\!-0.2$} and \hbox{$\Delta\!=\!0.8$} case are given,
respectively, by
\hbox{$J_{{\rm l},b\,({\rm cr})}^{({\rm TD},{\rm NCFR})}(36)\!=\!0.2185(5)$}
and
\hbox{$J_{{\rm l},b\,({\rm cr})}^{({\rm NCFR},XY)}(36)\!=\!1.1675(5)$}.
We have performed these DMRG calculations for various $\Delta$'s with
$J_{{\rm l},a}$ fixed at \hbox{$J_{{\rm l},a}\!=\!-0.2$}, and obtained
the phase transition line shown by the blue line in
Fig.~\ref{fig:phase-diagram}(a), supposing that the results in the
\hbox{$L\!=\!36$} system give good approximate results in the
\hbox{$L\!\to\!\infty$} limit~\cite{comment-2}.   Similarly, the phase
transition line shown by the blue line in Fig.~\ref{fig:phase-diagram}(b,c)
has been obtained by means of the DMRG calculations in the case of
\hbox{$J_{{\rm l},a}\!=\!0.2$}.

Figure~\ref{fig:GS-magnetization-etc}(a) suggests that the phase transition
between the TD and NCFR phases is of the second order, while that between the
NCFR and $XY$ phases is of the first order.  However, it is fairly
difficult to clarify the order of the phase transition by using only the
results of DMRG calculations.

We have also calculated the ground-state site magnetization
$m_{j,\ell;{\rm g}}(L;{\rm obc})$ by use of the DMRG
method~\cite{dmrg-white-1,dmrg-white-2}.  This quantity is defined by
\hbox{$m_{j,\ell;{\rm g}}(L;{\rm obc})\!=\!\langle S_{j,\ell}^z
\rangle_{L;{\rm g}}$}, where $\langle \cdots\rangle_{L;{\rm g}}$ denotes
the expectation value with respect to the ground state of the
Hamiltonian~(\ref{eq:hamiltonian}) under open boundary conditions.  Of
course, the relation \hbox{$\sum_{j=1}^L \big\{m_{j,a;{\rm g}}(L;{\rm obc})\!+
\!m_{j,b;{\rm g}}(L;{\rm obc})\bigr\}\!=\!M_{\rm g}(L;{\rm obc})$} holds.  In
Fig.~\ref{fig:GS-magnetization-etc}(b) we plot the $j$-dependences of
the ground-state rung magnetization
\hbox{$m_{j;{\rm g}}(L;{\rm obc})\bigl(=\!m_{j,a;{\rm g}}(L;{\rm obc})\!+
\!m_{j,b;{\rm g}}(L;{\rm obc})\bigr)$} and
$m_{j,\ell;{\rm g}}(L;{\rm obc})$, calculated for the \hbox{$L\!=\!36$}
system in the case where \hbox{$J_{{\rm l},a}\!=\!-0.2$},
\hbox{$J_{{\rm l},b}\!=\!0.7$} and \hbox{$\Delta\!=\!0.8$}; for these
parameters \hbox{$M_{\rm g}(L;{\rm obc})\!=\!15$}.  This figure demonstrates
that the $j$-dependences of these quantities are not uniform especially near
both of open boundaries.   Paying attention to this fact, we have examined the
Fourier transform $S_{q;{\rm g}}(L;{\rm obc})$ of
$m_{j;{\rm g}}(L;{\rm obc})$~\cite{hikihara-etal-1,hikihara-etal-2},
defined by 
\begin{equation}
    S_{q;{\rm g}}(L;{\rm obc}) = \frac{1}{\sqrt{L}}
        \sum_{j=1}^{L} \exp(iqj)\, 
           \biggl\{m_{j;{\rm g}}(L;{\rm obc})
                     -\frac{M_{\rm g}(L;{\rm obc})}{2L}\biggr\}\,,
\end{equation}
where $q$ is the wave number.  The squared modulus
$|S_{q;{\rm g}}(L;{\rm obc})|^2$ of this quantity, calculated for the
\hbox{$L\!=\!36$} system in the \hbox{$J_{{\rm l},a}\!=\!-0.2$},
\hbox{$J_{{\rm l},b}\!=\!0.7$} and \hbox{$\Delta\!=\!0.8$} case, where
\hbox{$M_{\rm g}(L;{\rm obc})\!=\!15$}, is plotted as a function of $q/\pi$ in
Fig.~\ref{fig:GS-magnetization-etc}(c).  This figure shows that the largest
peak of $|S_{q;{\rm g}}(36;{\rm obc})|^2$ appears at the position closest to
\hbox{$q=\pi$}, suggesting that the wave number of the dominant excitation in
the NCFR state is \hbox{$q=\pi$}.  $\bigl($Note that in the system with even
$L$ under
open boundary conditions, $|S_{q;{\rm g}}(L;{\rm obc})|^2$ at \hbox{$q=\pi$} is
exactly zero because of the space-inversion symmetry
\hbox{$m_{j;{\rm g}}(L;{\rm obc})=m_{L+1-j;{\rm g}}(L;{\rm obc})$}.$\bigr)$ We
therefore expect that the NCFR state has a commensurate character.  In order to
examine the commensurability of the NCFR state in full detail, it is necessary
to treat the Fourier transform of the rung magnetization
$m_{j;{\rm g}}(L;{\rm obc})$ as well as that of the ground-state two-spin
correlation function $\langle S_{j,\ell}^z\,S_{j',\ell}^z \rangle_{L;{\rm g}}$
in larger systems.   We will discuss this problem in the near future.

\section{Concluding remarks}

We have numerically determined, with the help of some physical considerations,
the ground-state phase diagrams of the \hbox{$S\!=\!1/2$} two-leg ladder with
different leg interactions, which is governed by the
Hamiltonian~(\ref{eq:hamiltonian}), in the cases where
\hbox{$J_{{\rm l},a}\!=\pm\,0.2$}, \hbox{$J_{\rm r}\!=\!-1$} and
\hbox{$0\!\leq\!\Delta\!<\!1$}.  The obtained phase diagrams on the $\Delta$
versus $J_{{\rm l},b}$ plane are shown in Fig.~\ref{fig:phase-diagram}.  The
characteristic features of the results are as follows:

\vspace{-0.28cm}

\begin{itemize}
\parsep=0pt
\itemsep=0pt
\parskip=0pt
\vspace{0.10cm}
\item[1)] The NCFR state appears as
the ground state in the region where
\hbox{$J_{{\rm l},a} J_{{\rm l},b}\!<\!0$}, when $\Delta$ is not too small.

\item[2)] The direct-product TD state is the {\it exact ground state}, when
\hbox{$J_{{\rm l},a}\!+\!J_{{\rm l},b}\!=\!0$} and
\hbox{$0\!\leq\!\Delta\lsim 0.83$}.

\end{itemize}
\vspace{-0.14cm}
It is emphasized that these results are attributed to the frustration effect.

We hope that the present research stimulates future experimental studies on
related subjects, which include the synthesization of spin ladder systems with
different leg interactions.


\ack

We would like to express our sincere thanks to Professors K Hida and
H Yamaguchi for their invaluable discussions and comments. This work has been
partly supported by JSPS KAKENHI Grant Numbers 15K05198,
16K05419 and 15K05882 (J-Physics) and also by Hyogo Science and Technology Association.   Finally, we thank the Supercomputer
Center, Institute for Solid State Physics, University of Tokyo and the
Computer Room, Yukawa Institute for Theoretical Physics, Kyoto University for
computational facilities.


\section*{References}
\begin{thebibliography}{9}

\bibitem{frustrated-leg-ladder1} Lavar{\'e}lo A, Guillaume G and Laflorencie N
2011 \PR B {\bf 84} 144407 and references therein

\bibitem{frustrated-leg-ladder2} Vekua T and Honecker A 2006 \PR B {\bf 73}
214427 and references therein

\bibitem{frustrated-leg-ladder3} Michaud F, Coletta T, Manmana S R, Picon J-D
and Mila F 2010 \PR B {\bf 81} 014407 and references therein

\bibitem{ladder-ral-tone} Tonegawa T, Okamoto K, Hikihara T and Sakai T 2016
{\it J. Phys.: Conf. Series} {\bf 683} 012039

\bibitem{inv-1} Okamoto K and Ichikawa Y 2002
{\it J. Phys. Chem. Solids} {\bf 63} 1575

\bibitem{inv-2} Okamoto K 2002 {\it Prog. Theor. Phys.} Suppl. No.145
208

\bibitem{inv-3} Tokuno A and Okamoto K 2005 {\it J. Phys. Soc. Jpn.}
{\bf 74} Suppl.~157

\bibitem{inv-4} Okamoto K 2014 {\it JPS Conf. Proc.} {\bf 1} 012031

\bibitem{ladder-ral-amiri} Amiri F, Sun G, Mikeska H-J and Vekua T 2015
\PR B {\bf 92} 184421

\bibitem{ladder-ral-japa} Japaridze G I and Pogosyan E 2006 \JPCM {\bf 18}
9297

\bibitem{OYA} Oshikawa M, Yamanaka M and Affleck I 1997 \PRL {\bf 78} 1984

\bibitem{tsukano-takahashi} Tsukano M and Takahashi M 1997
{\it J. Phys. Soc. Jpn.} {\bf 66} 1153

\bibitem{HTOS} Hikihara T, Tonegawa T, Okamoto K and Sakai T
in preparation

\bibitem{yamaguchi-etal-1} Yamaguchi H, Iwase K, Ono T, Shimokawa T,
Nakano H, Shimura Y, Kase N, Kittaka S, Sakakibara T, Kawakami T,
and Hosokoshi Y 2013 \PRL {\bf 110} 157205

\bibitem{yamaguchi-etal-2} Yamaguchi H, Miyagai H, Shimokawa T, Iwase K,
Ono T, Kono Y, Kase N, Araki K, Kittaka S, Sakakibara T, Kawakami T,
Okunishi K and Hosokoshi Y 2014 {\it J. Phys. Soc. Jpn.} {\bf 83} 033707

\bibitem{dmrg-white-1} White S R 1992 \PRL {\bf 69} 2863

\bibitem{dmrg-white-2} White S R 1993 \PR B {\bf 48} 10345

\bibitem{chen-etal} Chen W, Hida K and Sanctuary B C 2003 \PR B {\bf 67} 104401

\bibitem{hu-etal} Hu S, Normand B, Wang X and Yu L 2011 \PR B {\bf 84}
220402(R)

\bibitem{wierschem-etal} Wierschem K and Sengupta P 2014 \PR B {\bf 90} 115157
         
\bibitem{yoshikawa-miyashita} Yoshikawa S and Miyashita S 2005
{\it J. Phys. Soc. Jpn. Suppl.} {\bf 74} 71

\bibitem{BKT-1} Berezinskii Z L 1971 {\it Sov.\ Phys. JETP} {\bf 34} 610
                            
\bibitem{BKT-2} Kosterlitz J M and Thouless D J 1973 {\it J. Phys.} C {\bf 6}
1181

\bibitem{LSmethod-1} Okamoto K and Nomura K 1992 \PL A {\bf 169} 433

\bibitem{LSmethod-2} Nomura K and Okamoto K 1994 {\it J. Phys.} A {\bf 27} 5773

\bibitem{LSmethod-3} Kitazawa A 1997 {\it J. Phys.} A {\bf 30} L285

\bibitem{LSmethod-4} Nomura K and Kitazawa A 1998 {\it J. Phys.} {\bf 31} 7341

\bibitem{comment-1} We note that in the TD, $XY$ and H phase regions,
$E_0(L,0;{\rm pbc})$ always gives the ground-state energy of the finite-$L$
system under periodic boundary conditions.

\bibitem{comment-3} Here, to solve the equation \hbox{$A\!=\!B\!<C$} means that
to solve the equation \hbox{$A\!=\!B$} under the conditions \hbox{$A\!<\!C$}
and \hbox{$B\!<\!C$} (see Fig.~\ref{fig:excitation-energy}).

\bibitem{comment-2} We have made sure that, for a few values of $J_{{\rm l},a}$
and $\Delta$, the results in the \hbox{$L\!=\!48$} system are not so much
different from those in the \hbox{$L\!=\!36$} system.

\bibitem{hikihara-etal-1} Hikihara T, Kecke L, Momoi T and Furusaki A 2008
\PR B {\bf 78} 144404

\bibitem{hikihara-etal-2} Hikihara T, Momoi T, Furusaki A and Kawamura H 2010
\PR B {\bf 81} 224433

\end{thebibliography}
\end{document}